\documentclass[aps,prd,preprint,showpacs,nofootinbib,superscriptaddress]{revtex4-1}
\usepackage{amsmath}
\usepackage[utf8]{inputenc}
\usepackage{latexsym}
\usepackage{amsfonts}
\usepackage{graphicx,color}
\usepackage{mathrsfs}
\usepackage{epstopdf}
\usepackage{subfigure}
\usepackage{hyperref}\hypersetup{pdftitle=aa}

\begin{document}
\title{Inflation with Gauss-Bonnet coupling}

\author{Zhu Yi}
\email{yizhu92@hust.edu.cn}
\affiliation{School of Physics, Huazhong University of Science and Technology,
Wuhan, Hubei 430074, China}
\author{Yungui Gong}
\email{Corresponding author. yggong@mail.hust.edu.cn}
\affiliation{School of Physics, Huazhong University of Science and Technology,
Wuhan, Hubei 430074, China}
\author{Mudassar Sabir}
\email{msabir@hust.edu.cn}
\affiliation{School of Physics, Huazhong University of Science and Technology,
Wuhan, Hubei 430074, China}
\date{\today}

\begin{abstract}
We consider inflationary models with the inflaton coupled to the  Gauss-Bonnet term assuming a special relation $\delta_1=2\lambda\epsilon_1$
between the two slow-roll parameters $\delta_1$ and $\epsilon_1$. For the slow-roll inflation,
the assumed relation leads to the reciprocal relation between the Gauss-Bonnet coupling
function $\xi(\phi)$ and the potential $V(\phi)$,
and it leads to the relation $r=16(1-\lambda)\epsilon_1$ that reduces the tensor-to-scalar ratio $r$ by a factor of $1-\lambda$.
For the constant-roll inflation, we derive the analytical expressions
for the scalar and tensor power spectra, the scalar and tensor spectral tilts,  and the tensor-to-scalar ratio
to the first order of $\epsilon_1$ by using the method of Bessel function approximation.
The tensor-to-scalar ratio is reduced by a factor of $1-\lambda+\lambda\tilde \eta$.
Comparing the derived $n_s$-$r$ with the observations,
we obtain the constraints on the model parameters $\tilde\eta$ and $\lambda$.
\end{abstract}
\maketitle
\section{introduction}
The flatness and horizon problems in standard cosmology can be solved by cosmic inflation \cite{Guth:1980zm,Linde:1981mu,Albrecht:1982wi,Starobinsky:1980te,Sato:1980yn},
and the seeds of the large scale structure of our Universe
are sowed by the quantum fluctuations of the inflaton during inflation that leave imprints on the cosmic microwave background radiation
 \cite{Mukhanov:1981xt,Guth:1982ec,Hawking:1982cz,Bardeen:1983qw,Mukhanov:1985rz,Sasaki:1986hm}. The simplest inflation model is a canonical scalar filed with a flat potential minimally coupled to gravity.
 Since the current observations cannot tell the nature of the scalar field,
 there exist many other kinds of inflation models, and one of them is the Gauss-Bonnet inflation.
 The Gauss-Bonnet term that is induced from the superstring theory provides the possibility of
 avoiding the singularity problem of the Universe \cite{Antoniadis:1993jc,Kawai:1998ab,Kawai:1999pw,Tsujikawa:2001ad,Toporensky:2002ta}.
 Gauss-Bonnet coupling is also a subclass of the Horndeski theory in which
 equations of motion are, at most, of the second order in the derivative of both the
 metric $g_{\mu\nu}$ and the scalar field $\phi$ in four dimensions \cite{Horndeski:1974wa,Kobayashi:2011nu}.
The inflation models with the Gauss-Bonnet coupling have been studied in
 \cite{Guo:2009uk,Guo:2010jr,Jiang:2013gza,Koh:2014bka,Koh:2016abf,Kanti:2015pda,Hikmawan:2015rze,
Bhattacharjee:2016ohe,Wu:2017joj,Hikmawan:2017vno,Fomin:2017sqz,Mathew:2016anx,Granda:2017oku,
vandeBruck:2017voa,Kanti:1998jd,Nojiri:2005vv,Nojiri:2005jg,Nozari:2017rta,Rizos:1993rt,Heydari-Fard:2016nlj,Odintsov:2018zhw,Satoh:2008ck,Satoh:2007gn,Sberna:2017xqv}.
Among them, in Refs. \cite{Guo:2009uk,Guo:2010jr,Jiang:2013gza,Koh:2014bka},
the authors calculated the scalar tilt $n_s$ and the tensor-to-scalar ratio $r$ under the slow-roll condition $\epsilon_i\ll1$ and $\delta_i\ll1$.
The authors studied two special models with $V(\phi)=V_0\exp(-p \phi)$,
$\xi(\phi)=\xi_0\exp(p \phi)$ and $V(\phi)=V_0 \phi^p$, $\xi(\phi)=\xi_0 \phi^{-p}$, respectively,
which satisfy $V(\phi)\xi(\phi)=\text{const}$, and they find  the tensor-to-scalar ratio $r$ can be reduced.
Is the reduction of $r$ a generic feature of Gauss-Bonnet coupling or just an accidental effect of the specific potentials
and couplings?
In this paper,  we study this problem and show that for an arbitrary potential $V(\phi)$,
during slow-roll inflation if we choose the coupling function
$\xi(\phi)$ to satisfy the relation $V(\phi)\xi(\phi)=$ const,
then the tensor-to-scalar ratio $r$ is reduced. The reduction of $r$ brings the model to be
consistent with the  observations \cite{Ade:2015lrj,Array:2015xqh}. This provides  another mechanism to lower  the tensor-to-scalar ratio $r$ like the inflationary models with nonminimal derivative coupling\cite{Germani:2010gm,Yang:2015pga}. We also show that under the slow-roll approximation, the reciprocal relation between the coupling function $\xi(\phi)$
and the potential $V(\phi)$ can be derived from the relation
\begin{equation}
\label{condition1}
\delta_1=2\lambda \epsilon_1,
\end{equation}
where $\lambda$ is an order-one constant.

Besides the slow-roll inflationary scenario  there exists  a constant-roll inflationary scenario
\cite{Tsamis:2003px,Kinney:2005vj,Martin:2012pe,Motohashi:2014ppa,
Motohashi:2017vdc,Motohashi:2017aob,Nojiri:2017qvx,Oikonomou:2017bjx,
Odintsov:2017qpp,Dimopoulos:2017ged,Gao:2017owg,Ito:2017bnn,Karam:2017rpw,
Cicciarella:2017nls,Anguelova:2017djf,Gao:2018tdb,Gao:2018cpp,Yi:2017mxs,Mohammadi:2018wfk,Morse:2018kda,GalvezGhersi:2018haa,Pattison:2018bct}
in which one of the slow-roll parameter is regarded as constant instead of small and the slow-roll condition may be violated. The
constant-roll inflation has a richer physics than the slow-roll inflation does. For example, it can generate large local non-Gaussianity and the curvature perturbation may grow on the superhorizon scales
\cite{Martin:2012pe,Motohashi:2014ppa,Namjoo:2012aa}.
Furthermore, it can be used to generate the primordial black holes \cite{Germani:2017bcs,Gong:2017qlj,Motohashi:2017kbs}.
In this paper, we study the constant-roll inflation with the Gauss-Bonnet coupling, with the assumed relation \eqref{condition1}.
For the canonical constant-roll inflation with $\eta_H$ being a constant \cite{Yi:2017mxs},
the model is consistent with the observations only at the $2\sigma$ C.L.
With the help of Gauss-Bonnet coupling and the condition \eqref{condition1}, the model with constant $\eta_H$
is consistent with the observations at the $1\sigma$ C.L..
To discuss more general cases,  we introduce the slow-roll parameter $\tilde{\eta}_A=\epsilon_2-A \epsilon_1$ with a constant $A$,
and assume $\tilde{\eta}_A$ to be a constant.  For $A=0$, we have $\tilde{\eta}_0=\epsilon_2$; and for $A=2$, we have $\tilde{\eta}_2=-2\eta_H$,
so a different constant-roll model corresponds to a different choice of the value of $A$.

This paper is organized as follows. In Secs. \ref{sec-2a} and \ref{sec-2b}, we briefly review the slow-roll Gauss-Bonnet inflation.
In Sec. \ref{sec-3}, we show that under the slow-roll condition,
the relation $\xi(\phi)V(\phi)=\text{const}$ can be derived from the condition \eqref{condition1}.
We also discuss the effects of the Gauss-Bonnet coupling on the natural inflation and the $\alpha$-attractor with the
condition \eqref{condition1}. In  Sec. \ref{sec-4}, we study the constant-roll inflation models with the Gauss-Bonnet coupling under the condition \eqref{condition1}.
The paper is concluded in Sec. \ref{sec-5}.

\section{The slow-roll Gauss-Bonnet inflation}
\label{sec-2}
\subsection{The background}
\label{sec-2a}

In this section, we review the slow-roll inflation with the Gauss-Bonnet coupling. The action for the Gauss-Bonnet inflation is
\begin{equation}\label{action1}
  S=\frac{1}{2}\int\sqrt{-g}d^4 x \left[R-g^{\mu\nu}\partial_\mu\phi\partial_\nu\phi-2V(\phi)-\xi(\phi)R_{GB}^2\right],
\end{equation}
where $R^{2}_{\rm GB}
= R_{\mu\nu\rho\sigma} R^{\mu\nu\rho\sigma} - 4 R_{\mu\nu}
R^{\mu\nu} + R^2$ is the Gauss-Bonnet term which is a pure topological term in four dimensions,
and $\xi(\phi)$ is the Gauss-Bonnet coupling function.
With the Friedmann-Robertson-Walker metric, the field equations are
\begin{gather}
\label{beq1}
 6H^2 = \dot{\phi}^2 + 2V + 24\dot{\xi}H^3,\\
 \label{beq12}
 2\dot{H} = -\dot{\phi}^2 + 4\ddot{\xi}H^2 +
4\dot{\xi} H \left(2\dot{H} - H^2\right), \\
 \left(\ddot{\phi} + 3 H \dot{\phi}\right) + V_{,\phi} +12\xi_{,\phi}H^2 \left(\dot{H}+H^2\right) = 0.
\label{beq2}
\end{gather}
where  a dot  denotes the derivative with respect to time $t$, and $V_{,\phi}=dV/d\phi$.

For the slow-roll inflation, we introduce the following slow-roll conditions
\begin{equation}\label{con:sl1}
  \dot{\phi}^2\ll V(\phi),\quad |\ddot{\phi}|\ll 3H|\dot{\phi}|, \quad 4 H|\dot{\xi}|\ll1,\quad |\ddot{\xi}|\ll H|\dot{\xi}|.
\end{equation}
Under these slow-roll conditions, Eqs. \eqref{beq1}, \eqref{beq12}, and \eqref{beq2} become
\begin{gather}
\label{seq1}
 H^2 \approx \frac13 V, \\
\label{seq12}
\dot{H} \approx -\frac12   \dot{\phi}^2 - 2\dot{\xi}H^3, \\
\label{seq2}
\dot{\phi} \approx -\frac{1}{3 H}(V_{,\phi} + 12\xi_{,\phi}H^4).
\end{gather}
To quantify the slow-roll conditions, we introduce the hierarchy of Hubble flow parameters \cite{Schwarz:2001vv},
\begin{equation}\label{sl1}
  \epsilon_1=-\frac{\dot{H}}{H^2},\quad \epsilon_{i+1}=\frac{d\ln|\epsilon_i|}{d\ln a},\quad i\geq1,
\end{equation}
and the hierarchy of the flow parameters for the coupling function \cite{Guo:2010jr}
\begin{equation}\label{sxi1}
  \delta_1=4\dot{\xi}H,\quad \delta_{i+1}=\frac{d\ln|\delta_i|}{d\ln a},\quad i\geq 1.
\end{equation}
In terms of these slow-roll parameters,  the  slow-roll conditions \eqref{con:sl1} become
\begin{equation}
\label{sl12}
\epsilon_1\ll1,\quad |\epsilon_2|\ll1,\quad
|\delta_1|\ll1,\quad |\delta_2|\ll1.
\end{equation}
With the help of  Eqs. \eqref{seq1}, \eqref{seq12}, and \eqref{seq2}, the slow-roll parameters can  be expressed by the potential $V(\phi)$ and the  coupling function $\xi(\phi)$ as
\begin{gather}
\label{eps1}
\epsilon_1  \approx  \frac{Q}{2} \frac{V_{,\phi}}{V},\\
\label{eps2}
\epsilon_2  \approx  -Q \left(\frac{V_{,\phi\phi}}{V_{,\phi}}
 - \frac{V_{,\phi}}{V} + \frac{Q_{,\phi}}{Q}\right), \\
 \label{del1}
\delta_1  \approx -\frac43 \xi_{,\phi} Q V ,\\
 \label{del2}
 \delta_2 \approx -Q \left(\frac{\xi_{,\phi\phi}}{\xi_{,\phi}} +
\frac{V_{,\phi}}{V}
 + \frac{Q_{,\phi}}{Q}\right),
\end{gather}
where $Q =  V_{,\phi}/V + 4\xi_{,\phi}V/3$. The $e$-folding number $N$ at the horizon exit before the end of the inflation can also be expressed by the potential and the  coupling function
\begin{equation}
\label{ne}
N(\phi) \approx \int_{\phi_{e}}^{\phi}
 \frac{3  V}{3 V_{,\phi}+4\xi_{,\phi}V^2}d\phi=\int_{\phi_{  e }}^{\phi}\frac{d\phi}{Q}.
\end{equation}
\subsection{The power spectrum}
\label{sec-2b}

\subsubsection{The scalar perturbation}
In the flat gauge $\delta\phi=0$, the gauge invariant scalar perturbation becomes the curvature perturbation
which is related to the  metric perturbation by $\delta g_{ij}=a^2(1+2\zeta)\delta_{ij}$.
The Fourier component of the mode function $v_k=z_s\zeta_k$ for the curvature perturbation $\zeta$ satisfies the Mukhanov-Sasaki equation \cite{Mukhanov:1985rz,Sasaki:1986hm,Hwang:1999gf,Cartier:2001is,Hwang:2005hb,DeFelice:2011zh}
\begin{equation}\label{ms1}
  v_k'' + \left(c_{s}^2 k^2 - \frac{z''_{s}}{z_{s}}\right)v_k = 0,
\end{equation}
where a prime represents the derivative with respect to the conformal time $\tau=\int a^{-1} dt$. The sound speed $c_s$ and $z_s$ are
\begin{gather}
 c_{s}^2 = 1 - \Delta^2 \frac{2\epsilon_1+\frac12 \delta_1(1-5\epsilon_1-\delta_2)}{F},\\
 z_{s}^2 = a^2 \frac{F}{(1-\frac12 \Delta)^2},
\end{gather}
where $\Delta = \delta_1/(1-\delta_1)$, $F=
2\epsilon_1-\delta_1(1+\epsilon_1-\delta_2)+3\Delta \delta_1/2$, and
\begin{eqnarray}
\label{zrpp}
\frac{z_{s}''}{z_{s}} &=& a^2H^2 \Bigg[2 - \epsilon_1
 +\frac32 \frac{\dot{F}}{H F}
 +\frac32 \frac{\dot{\Delta}}{H(1-\frac12 \Delta)} \nonumber \\
&&
 +\frac12 \frac{\ddot{F}}{H^2F}
 +\frac12 \frac{\ddot{\Delta}}{H^2(1-\frac12 \Delta)}
 -\frac14 \frac{\dot{F}^2}{H^2F^2} \nonumber \\
&&
 +\frac12 \frac{\dot{\Delta}^2}{H^2(1-\frac12 \Delta)^2}
 +\frac12 \frac{\dot{\Delta}}{H(1-\frac12 \Delta)}\frac{\dot{F}}{H F}
 \Bigg]\nonumber \\
 &&=\frac{1}{\tau^2}\left(\nu^2-\frac14\right).
\end{eqnarray}
With the slow-roll conditions \eqref{con:sl1}, we get
\begin{equation}\label{slah}
  aH\approx -\frac{1}{(1-\epsilon_1)\tau}.
\end{equation}
Substituting Eq. \eqref{slah} into Eq. \eqref{zrpp}, we obtain
\begin{equation}\label{slnu}
  \nu=\frac32+\epsilon_1+\frac{2\epsilon_1\epsilon_2-\delta_1\delta_2}{2(2\epsilon_1-\delta_1)}.
\end{equation}
Assuming that $\nu$ is almost a constant,  we  get the power spectrum for the  scalar perturbation expressed by the Hankel function
\begin{equation}\label{slpr0}
\mathcal{P_R}=\frac{k^3}{2\pi^2}\left|\zeta_k\right|^2=\frac{H^2}{8\pi}
\frac{(1-\Delta/2)^2}{(1-\epsilon_1)Fc_s^3}\left[H_{\nu}^{(1)}
\left(\frac{1}{1-\epsilon_1}\frac{c_sk}{aH}\right)\right]^2\left(\frac{c_s k}{aH}\right)^3.
\end{equation}
On superhorizon scales, $c_s k\ll aH$, using the asymptotic behavior of the Hankel function, the power spectrum for the scalar perturbation becomes
\begin{equation}\label{slpr1}
\mathcal{P_R}=2^{2\nu-3}\left[\frac{\Gamma(\nu)}{\Gamma(3/2)}\right]^2
\frac{(1-\Delta/2)^2}{Fc_s^3}\left(\frac{H}{2\pi}\right)^2
\left(1-\epsilon_1\right)^{2\nu-1}\left.\left(\frac{c_s k}{aH}\right)^{3-2\nu}\right|_{c_sk=aH}.
\end{equation}
Therefore,  the scalar spectral tilt is \cite{Guo:2010jr}
\begin{equation}\label{ns}
\begin{aligned}
  n_s-1=\frac{d\ln\mathcal{P_R}}{d \ln  k}&=3-2\nu\\
  &=-2\epsilon_1 -\frac{2\epsilon_1\epsilon_2-\delta_1\delta_2}{2\epsilon_1-\delta_1}.
\end{aligned}
\end{equation}

\subsubsection{The tensor perturbation}
For the tensor perturbation $\delta g_{ij} = a^2h_{ij}$, the mode function $u^\lambda_k(\tau)=z_T h^\lambda_k/2$  satisfies the equation \cite{Hwang:1999gf,Cartier:2001is,Hwang:2005hb,DeFelice:2011zh}
\begin{equation}\label{ms2}
  \frac{d^2u^\lambda_k}{d\tau^2} + \left(c_{T}^2 k^2 - \frac{z''_{T}}{z_{T}}\right)u^\lambda_k = 0,
\end{equation}
where ``$\lambda$" stands for the ``$+$" or ``$\times$" polarizations and
\begin{eqnarray}
&& z_{T}^2 = a^2 (1-\delta_1), \\
&& c_{T}^2 = 1 + \Delta (1-\epsilon_1-\delta_2).
\end{eqnarray}
In terms of the  slow-roll parameters, we have
\begin{equation}
\begin{aligned}
\label{ztpp}
\frac{z_{T}''}{z_{T}} &=a^2H^2
 \bigg[2-\epsilon_1-\frac32 \Delta\delta_2-\frac12\Delta\delta_2(-\epsilon_1+\delta_2+\delta_3)
&-\frac14\Delta^2\delta_2^2 \bigg]\\
 &=\frac{1}{\tau^2}\left(\mu^2-\frac14\right).
\end{aligned}
\end{equation}
By using the slow-roll conditions \eqref{con:sl1} and with the help of Eq. \eqref{slah}, we get
\begin{equation}\label{slmu}
  \mu\approx\frac32+\epsilon_1.
\end{equation}
Assuming that $\mu$ is almost a constant, following the same procedure as that in scalar perturbation, we obtain the power spectrum for the  tensor perturbation
\begin{equation}\label{slpt0}
  \mathcal{P_T}=\frac{k^3}{2\pi^2}\sum_{\lambda=+,\times}\left|\frac{2u^\lambda_k}{z_T}\right|^2
  =\frac{H^2}{\pi(1-\epsilon_1)(1-\delta_1)c_T^3}
  \left[H^{(1)}_{\mu}\left(\frac{1}{1-\epsilon_1}\frac{c_T k}{aH}\right)\right]^2\left(\frac{c_T k}{aH}\right)^3.
\end{equation}
On superhorizon scales, $c_T k\ll aH$, we have
\begin{equation}\label{slpt1}
  \mathcal{P_T}=\frac{2^{2\mu}}{(1-\delta_1)c_T^3}\left[\frac{\Gamma(\mu)}{\Gamma(3/2)}\right]^2
  \left(\frac{H}{2\pi}\right)^2\left(1-\epsilon_1\right)^{2\mu-1}\left.\left(\frac{c_Tk}{aH}\right)^{3-2\mu}\right|_{c_Tk=aH}.
\end{equation}
The tensor spectral tilt is \cite{Guo:2010jr}
\begin{equation}\label{slnT}
  n_T=\frac{d\ln\mathcal{P_T}}{d \ln k}=3-2\mu=-2\epsilon_1,
\end{equation}
and the tensor-to-scalar ratio is \cite{Guo:2010jr}
\begin{equation}\label{r}
  r=\frac{\mathcal{P_T}}{\mathcal{P_R}}=16\epsilon_1-8\delta_1.
\end{equation}
The time  at the horizon crossing, $aH=c_s k$, for the scalar perturbation is not exactly the same as that for the tensor perturbation,
$aH=c_T k$, however to the lowest order of the slow-roll approximation, this difference is unimportant \cite{Guo:2010jr}.
\subsection{The models}
\label{sec-3}
In \cite{Jiang:2013gza}, the authors studied two special models with $V(\phi)=V_0\exp(-p \phi)$,
$\xi(\phi)=\xi_0\exp(p \phi)$ and $V(\phi)=V_0 \phi^p$, $\xi(\phi)=\xi_0 \phi^{-p}$, respectively.
The potentials and coupling functions in the two models satisfy the relation $V(\phi) \xi(\phi)=$ const.
Now we  show that the reciprocal relation between the coupling
function $\xi(\phi)$ and the potential $V(\phi)$ can be obtained from the condition
\eqref{condition1} under the slow-roll conditions \eqref{con:sl1}.
Substituting Eqs. \eqref{eps1} and \eqref{del1} into Eq. \eqref{condition1}, and choosing the integration constant to be zero, we get
\begin{equation}\label{xiphi}
  \xi(\phi)=\frac{3\lambda}{4 V(\phi)}.
\end{equation}
As pointed out in Ref. \cite{vandeBruck:2016xvt},
while the potential becomes smaller during inflation, the Gauss-Bonnet coupling term grows due to the relation \eqref{xiphi},
and this implies a slow-down of inflation and the reheating does not happen.
To avoid the reheating problem, following Ref. \cite{vandeBruck:2016xvt}, we introduce a small energy parameter $\Lambda_0$ and instead use the relation
\begin{equation}\label{xiphi2}
    \xi(\phi)=\frac{3\lambda}{4 V(\phi)+\Lambda_0},
\end{equation}
where $\Lambda_0\ll (10^{16}\text{Gev})^4$. During  the inflationary phase, $\Lambda_0$ is very small compared to the potential $V(\phi)$,
and it can be ignored, Eq. \eqref{xiphi2} reduces to Eq. \eqref{xiphi}. At the reheating phase, when $V(\phi)\approx 0$, $\Lambda_0$ becomes important
and it will regulate $\xi(\phi)$ to prevent it from diverging. Thereby it effectively avoids the reheating problem.
Since we observe about $16$ $e$-folds of inflation only, Eq. \eqref{xiphi} remains a good approximation to Eq.
\eqref{xiphi2} over the observable range, so the condition \eqref{condition1} is valid over the observable range.
Therefore, we use the coupling function \eqref{xiphi2} for the model and the approximation \eqref{xiphi} far away from
the end of inflation during which the calculation is carried on in this paper.

By using the condition \eqref{condition1}, and the definitions \eqref{sl1} and \eqref{sxi1}, we obtain the relations for the other slow-roll parameters
\begin{equation}\label{conditioni}
\delta_{i+1}= \epsilon_{i+1},\quad i \geq 1.
\end{equation}
Substituting the relations \eqref{condition1} and \eqref{conditioni} into Eqs. \eqref{ns} and \eqref{r}, we obtain
\begin{gather}
\label{ns1}
n_s-1= -2\epsilon_1-\epsilon_2,\\
\label{r1}
r=16(1-\lambda)\epsilon_1.
\end{gather}
The result for the scalar spectral tilt $n_s$ is the same as that in the canonical case without the Gauss-Bonnet coupling, but the result for the tensor-to-scalar ratio $r$ is reduced by a factor of $1-\lambda$ comparing with the canonical  case without the Gauss-Bonnet coupling. To obtain these results we only assumed the slow roll conditions without any further constraints on the form of the potential. The reduction of the tensor-to-scalar ratio $r$ due to the Gauss-Bonnet coupling
and the condition \eqref{condition1} can help more inflationary models like the chaotic inflation and natural inflation
to be consistent with the observations, this is the main motivation for the condition \eqref{condition1}.

In the canonical case without the Gauss-Bonnet coupling, we have the Hubble flow and horizon flow slow-roll parameters,
and different definitions give the same result.
In the case with the Gauss-Bonnet coupling, different definitions may give different results.
For comparison, we introduce the following slow-roll parameters
\begin{equation}
\label{etapa}
\eta_{H}=-\frac{\ddot{H}}{2 H\dot{H}},\quad
\eta_{H\phi}=\frac{2H_{,\phi\phi}}{H},\quad
\eta_{\phi}=-\frac{\ddot{\phi}}{H\dot{\phi}},\quad \eta_{\xi}=\frac{\ddot{\xi}}{H\dot{\xi}}.
\end{equation}
Furthermore, we also introduce the potential slow-roll parameters
\begin{equation}
\label{pslp}
\epsilon_{V}=\frac12\left(\frac{V_{,\phi}}{V}\right)^2,\quad \eta_{V}=\frac{V_{,\phi\phi}}{V}.
\end{equation}
By using the condition \eqref{condition1}, and the slow-roll conditions \eqref{con:sl1},
to the first order of approximation, we obtain the relations
\begin{gather}
\label{epv1}
\epsilon_V\approx \frac{\epsilon_1}{1-\lambda},\\
\label{etah}
\eta_{H}=-\frac12(\epsilon_2-2\epsilon_1)\approx\eta_{\phi},\\
\label{etahphi1}
\eta_{H\phi}\approx-\frac{1}{2(1-\lambda)}(\epsilon_2-2\epsilon_1), \\
\label{etaxi1}
\eta_{\xi}=\epsilon_2+\epsilon_1,\\
\label{etav1}
\eta_{V}\approx-\frac{1}{2(1-\lambda)}(\epsilon_2-4\epsilon_1).
\end{gather}
These relations can be parametrized as
\begin{equation}\label{slpara}
 \tilde{\eta}_A=\epsilon_2-A\epsilon_1,
\end{equation}
where $A$ is a constant. For $A=0$, we get $\tilde{\eta}_0=\epsilon_2$. For $A=2$, we get $\tilde{\eta}_2=-2\eta_H$. For $A=4$,
we get $\tilde{\eta}_4=-2(1-\lambda)\eta_V$ under the slow-roll conditions.
Now we apply the above results to some specific models.

\subsubsection{Natural inflation}
For the natural inflation
\begin{equation}\label{natural}
  V=V_0\left[1+\cos\left(\frac{\phi}{f}\right)\right],
\end{equation}
the potential slow-roll parameters are
\begin{gather}\label{naturalsl1}
  \epsilon_V=\frac{\sin ^2\left(\phi/f\right)}{2 f^2 \left[\cos \left(\phi/f\right)+1\right]^2}, \\
  \label{naturalsl2}
  \eta_V=-\frac{\cos \left(\phi/f\right)}{f^2 \left[\cos \left(\phi/f\right)+1\right]}.
\end{gather}

\begin{figure}
  \centering
  \includegraphics[width=0.5\textwidth]{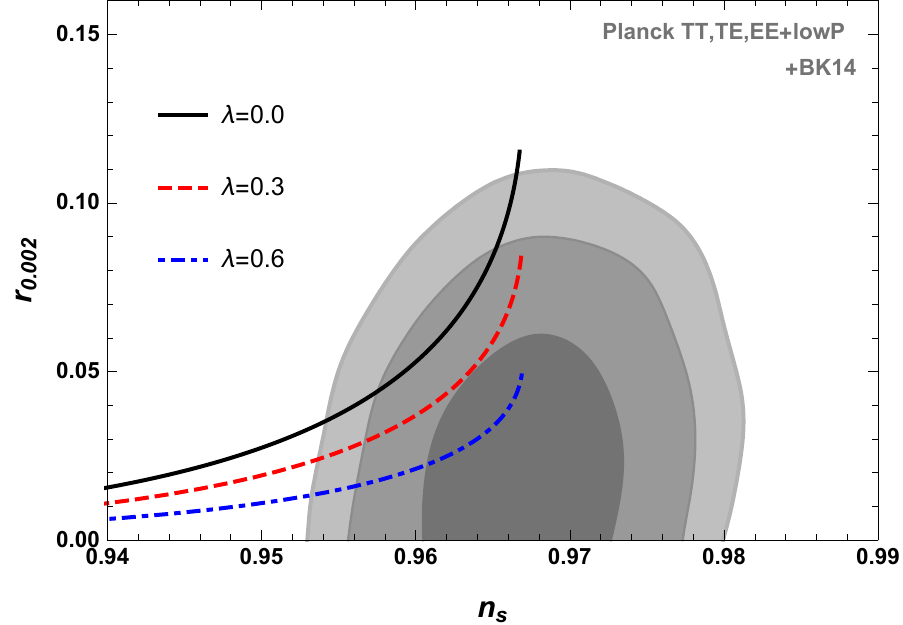}
  \caption{The marginalized $1\sigma$, $2\sigma$ and $3\sigma$ confidence level contours for $n_s$ and $r$ from Planck 2015 and BICEP2/Keck data \cite{Ade:2015lrj,Array:2015xqh} along with the observational constraints on $n_s$-$r$ for
  the natural inflation with different values of $\lambda$ and $f$. The solid black curve, the dashed red curve and the dashdotted blue curve represent the results for $\lambda=0$, $\lambda=0.3$ and $\lambda=0.6$, respectively.}\label{p1}
\end{figure}

The value of the inflaton at the end of inflation is
\begin{equation}\label{naturalphie}
\phi_e=f \arccos\left(\frac{1-2\tilde{f}^2}{1+2 \tilde{f}^2}\right),
\end{equation}
where $\tilde{f}=f/\sqrt{1-\lambda}$.
The $e$-folding number $N$ at the horizon exit for a pivotal scale $k_*$ is
\begin{equation}\label{naturalphii}
  N=2\tilde{f}^2\ln\left[\frac{\sin\left(\phi_e/2f\right)}
  {\sin\left(\phi_*/2f\right)}\right].
\end{equation}
The  scalar spectral tilt and the tensor-to-scalar ratio $r$ are
\begin{gather}
\label{natrual:ns}
  n_s-1=-\frac{1}{\tilde{f}^2} \frac{1+\exp[- \tilde{N}/\tilde{f}^2]}{1-\exp[-\tilde{N}/\tilde{f}^2]},\\
\label{natrual:r}
  r=\frac{8(1-\lambda) }{\tilde{f}^2} \frac{\exp[- \tilde{N}/\tilde{f}^2]}{1-\exp[- \tilde{N}/\tilde{f}^2]},
\end{gather}
where
\begin{equation}\label{natural:N}
  \tilde{N}=N-\tilde{f}^2\ln\frac{2\tilde{f}^2}{1+2\tilde{f}^2}.
\end{equation}
We compare the predictions from Eqs. \eqref{natrual:ns} and \eqref{natrual:r}
for different values of $f$ and $\lambda$ with the observations \cite{Ade:2015lrj,Array:2015xqh}
and the results are displayed in Fig. \ref{p1}.
For the natural inflation without the Gauss-Bonnet coupling, $\lambda=0$, the predictions for $n_s$ and $r$  are only consistent within the $2\sigma$ confidence level of the observations because of  the large tensor-to-scalar ratio $r$.
With the help of the  Gauss-Bonnet coupling, the natural inflation can be consistent with the observations
at the  $1\sigma$  confidence level if $\lambda$ is large enough, due to the reduction mechanism.
\subsubsection{$\alpha$-attractors }
\begin{figure}[htbp]
\centering
\includegraphics[width=0.45\textwidth]{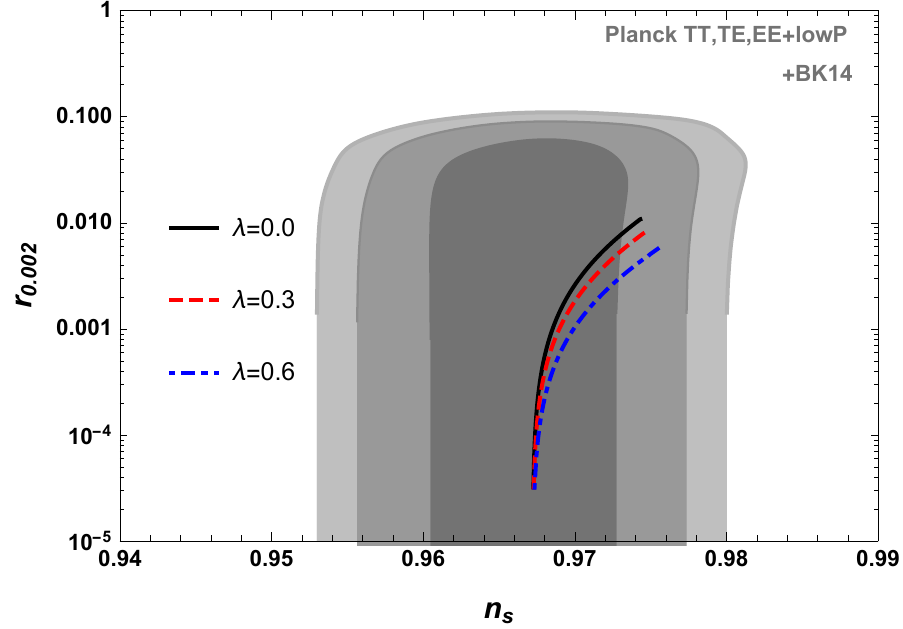}
\includegraphics[width=0.45\textwidth]{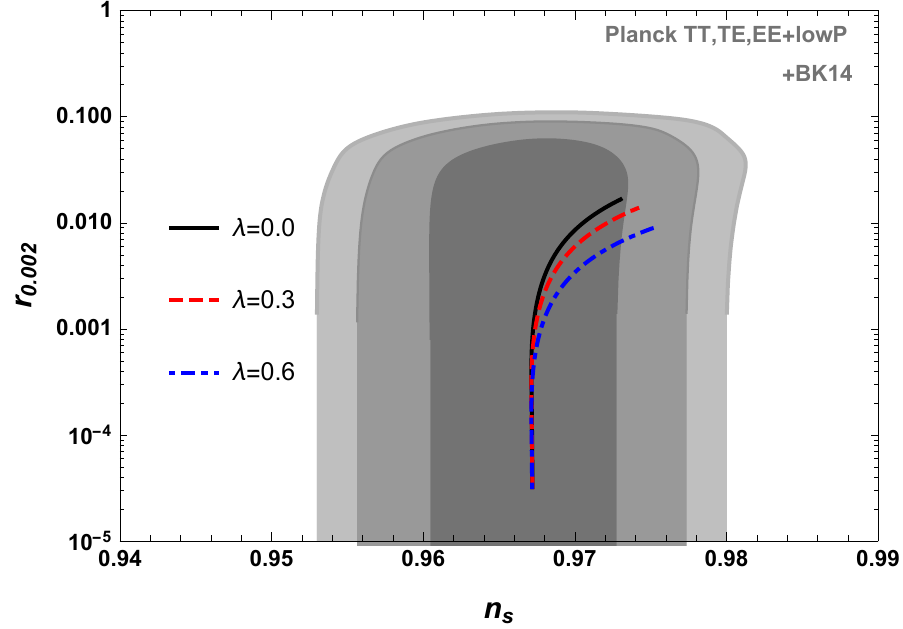}
\caption{Similar to Fig. \ref{p1}. The left panel shows the results for the E-model  $V=V_0\left[1-\exp\left(-\sqrt{2/ 3\alpha}\phi\right)\right]^{1/2}$,  and the right panel shows the results for the T-model $V(\phi)=V_0\tanh^{1/2}\left({\phi}/{\sqrt{6\alpha}}\right)$.}
\label{p23}
\end{figure}

For the E-model, the potential is
\begin{equation}\label{Emodel1}
  V=V_0\left[1-\exp\left(-\sqrt{\frac{2}{3\alpha}}\phi\right)\right]^{2n}.
\end{equation}
As an example, in this paper we  consider the case $n=1/4$ only. The scalar spectral tilt $n_s$ and the tensor-to-scalar ratio $r$ are  \cite{Yi:2016jqr}
\begin{gather}
\label{Emodel:ns}
n_s=1+\frac{2}{3 \tilde{\alpha} \left[ g(N,\tilde{\alpha})+1\right]}-\frac{5}{6\tilde{\alpha}  \left[ g(N,\tilde{\alpha})+1\right]^2},\\
\label{Emodel:r}
r=\frac{4 (1-\lambda)}{3 \tilde{\alpha} \left[g(N,\tilde{\alpha})+1\right]^2},
\end{gather}
where $\tilde{\alpha}=\alpha/(1-\lambda)$,
\begin{equation}
\label{Emodel2}
  g(N,\tilde{\alpha})=W_{-1}\left[-\left(\frac{1}{6 \tilde{\alpha} }+\frac{v}{6 \tilde{\alpha} }+1\right)
 \exp\left({-1-\frac{v+ 2N+1 }{6\tilde{\alpha }}}\right)\right],
\end{equation}
with $v=\sqrt{6\tilde{\alpha}+1}$, and $W_{-1}$ is the lower branch of the Lambert $W$ function. We compare the results  \eqref{Emodel:ns} and \eqref{Emodel:r} with the observations \cite{Ade:2015lrj,Array:2015xqh} and the comparisons are displayed in the left panel of Fig. \ref{p23}.

For the T-model, the potential is
\begin{equation}\label{Tmodel1}
 V(\phi)=V_0\tanh^{2n}\left(\frac{\phi}{\sqrt{6\alpha}}\right),
\end{equation}
Similar to the E-model inflation, we consider the special case $n=1/4$ only.
The  scalar spectral tilt $n_s$ and the tensor-to-scalar ratio $r$ are \cite{Yi:2016jqr}
\begin{gather}
\label{Tmodel:ns}
n_s=1-\frac2N+\frac{ (N+1) \sqrt{36 \tilde{\alpha} ^2+(1-6 \tilde{\alpha} )}-3 \tilde{\alpha} -(3 \tilde{\alpha} -2) N/2+1}{N \left[\left(\sqrt{9\tilde{\alpha} ^2+(1-6\tilde{\alpha} )/4}+N+1/2\right)^2-9 \tilde{\alpha}^2\right]},\\
\label{Tmodel:r}
r=\frac{12(1-\lambda)\tilde{\alpha}}{\left(\sqrt{9\tilde{\alpha}^2+(1-6\tilde{\alpha})/4}+ N+1/2\right)^2-9\tilde{\alpha}^2},
\end{gather}
where $\tilde{\alpha}=\alpha/(1-\lambda)$. We show the results along with the observational constraints in the right panel of Fig. \ref{p23}.

The $\alpha$-attractor models predict small tensor-to-scalar ratio $r$, with the help of the Gauss-Bonnet term, the tensor-to-scalar ratio $r$ becomes smaller.

\section{The constant-roll inflation}
\label{sec-4}

As discussed in the introduction, without the Gauss-Bonnet coupling, the constant-roll inflation with $\eta_H$ being a constant
is consistent with the observations only at the $2\sigma$ C.L. \cite{Yi:2017mxs}. Since the Gauss-Bonnet coupling with the condition \eqref{condition1} helps reducing the tensor-to-scalar ratio $r$,
so it is interesting to discuss the effect of the Gauss-Bonnet coupling on the constant-roll inflation.
In this section, we  study the constant-roll inflation by taking $\tilde{\eta}_A$ defined in Eq. \eqref{slpara} as a constant with the condition \eqref{condition1}.
Combining Eqs. \eqref{conditioni} and \eqref{slpara}, we  obtain
\begin{equation}\label{ddxi}
  4\ddot{\xi}=2\lambda\epsilon_1[(A+1)\epsilon_1+\tilde{\eta}_A].
\end{equation}
Substituting the result into the background Eqs. \eqref{beq1} and \eqref{beq12}, we  obtain
\begin{equation}\label{ephi}
  \epsilon_1=2\left(\frac{H_{,\phi}}{H}\right)^2\frac{1+\lambda(\tilde{\eta}_A-1)}{1-2\lambda(A-1)H_{,\phi}^2/H^2},
\end{equation}
and
\begin{equation}\label{eeta}
\eta_H=\frac{2H_{,\phi\phi}[1+\lambda(\tilde{\eta}_A-1)+\lambda(A-1)\epsilon_1]
 }{H}-\frac{\lambda(A-1)\tilde{\eta}_A\epsilon_1+\lambda A(A-1)\epsilon_1^2}{2[1+\lambda(\tilde{\eta}_A-1)+\lambda(A-1)\epsilon_1]}.
 \end{equation}
Combining Eqs. \eqref{ephi} and \eqref{eeta}, we get
\begin{equation}\label{eetaA}
  \tilde{\eta}_A=(2-A)\epsilon_1-\frac{4H_{,\phi\phi}[1+\lambda(\tilde{\eta}_A-1)+\lambda(A-1)\epsilon_1]
 }{H}+\frac{\lambda(A-1)\tilde{\eta}_A\epsilon_1+\lambda A(A-1)\epsilon_1^2}{1+\lambda(\tilde{\eta}_A-1)+\lambda(A-1)\epsilon_1}.
\end{equation}
where we used the relation $\epsilon_2=2\epsilon_1-2\eta_H$. Combining Eqs. \eqref{ephi} and \eqref{eetaA},
we may get the analytical form of $H(\phi)$ for constant $\tilde{\eta}_A$, and the potential can be obtained from
the Hamilton-Jacobi equation
\begin{equation}\label{hj}
  V(\phi)=(3-6\lambda\epsilon_1)H^2-2[1+\lambda(\tilde{\eta}_A-1)+\lambda(A-1)\epsilon_1]^2H_{,\phi}^2.
\end{equation}

Using the definition \eqref{sl1}, we obtain the solution to $\epsilon_1$ in terms of the $e$-folding number $N$,
\begin{equation}\label{const0e1}
  \epsilon_1(N)=\frac{\tilde{\eta}_A}{(A+\tilde{\eta}_A)\exp(\tilde{\eta}_A N)-A},
\end{equation}
where we used the relations $H dt=-dN$ and $\epsilon_1(0)=1$ at the end of inflation.
To find the relation between $aH$ and $\tau$, we use the relation
\begin{equation}
\label{tau:epsilon0}
\frac{d}{d\tau}\left(\frac{1}{aH}\right)=-1+\epsilon_1.
\end{equation}
Assuming that  $\epsilon_1$ is almost a constant, we get the following relation \footnote{The relation $(aH)^{-1}=[-1+\epsilon_1/(1-\tilde{\eta}_A)]\tau$ was derived in Ref. \cite{Yi:2017mxs}. However,
it is not applicable if $\epsilon_1$ is almost a constant.}
 \begin{equation}\label{const0ah}
  \frac{1}{aH}\approx \left(-1+\epsilon_1\right)\tau.
\end{equation}
Substitute Eq. \eqref{const0ah} into Eq. \eqref{zrpp}, we obtain
\begin{equation}\label{const0nu}
  \nu=\frac12|3+\tilde{\eta}_A|+\nu_{A}\epsilon_1,
\end{equation}
where
\begin{equation}\label{const0nu1}
\nu_{A}=\frac{a_0+a_1 \tilde{\eta}_A+a_2\tilde{\eta}_A^2+a_3\tilde{\eta}_A^3}{2|\tilde{\eta}_A+3|(1-\lambda+\lambda \tilde{\eta}_A)},
\end{equation}
and
\begin{equation}\label{cosnt0a}
\begin{aligned}
a_0&=3(2+A)(1-\lambda),\quad a_1= (5+2A)+4(1+A)\lambda+3\lambda^2,\\
a_2&=1+(6+4A)\lambda+8\lambda ^2,\quad a_3=(1+4\lambda)\lambda.
\end{aligned}
\end{equation}
Assuming that $\nu$ is almost a constant,  we derive the power spectrum for the scalar perturbation
\begin{equation}\label{const0pr0}
  \mathcal{P_{R}}=\frac{k^3}{2\pi^2}\left|\frac{v_k}{z_s}\right|^2=\frac{H^2}{8\pi}\frac{1}{(1-\epsilon_1)}\frac{(1-\Delta/2)^2}{c_s^3F}
\left[H_{\nu}^{(1)}\left(\frac{1}{1-\epsilon_1}\frac{c_s k}{aH}\right)\right]^2\left(\frac{c_sk}{aH}\right)^3.
\end{equation}
On superhorizon scales, $c_sk\ll aH$, using the asymptotic behavior of the Hankel function, the power spectrum for the scalar perturbation becomes
\begin{equation}\label{const0pr1}
  \mathcal{P_{R}}=\left.2^{2\nu-3}\left[\frac{\Gamma(\nu)}{\Gamma(3/2)}\right]^2\frac{(1-\Delta/2)^2}{c_s^3F}\left(\frac{H}{2\pi}\right)^2
  \left(1-\epsilon_1\right)^{2\nu-1}\left(\frac{c_sk}{aH}\right)^{3-2\nu}\right|_{aH=c_s k}.
\end{equation}
The expression is the same as that for the slow-roll inflation except that the value of $\nu$ is different.
The scalar spectral tilt is
\begin{equation}
\label{const0ns}
\begin{split}
n_s-1&=3-|3+\tilde{\eta}_A|-2\nu_A\times\epsilon_1\\
&=3-|3+\tilde{\eta}_A|-\frac{2\nu_A\tilde{\eta}_A}{(A+\tilde{\eta}_A)\exp(\tilde{\eta}_A N)-A}.
\end{split}
\end{equation}
Similarly, for the tensor perturbation, we obtain
\begin{equation}
\label{const0mu}
  \mu=\frac32+\left(1-\lambda\tilde{\eta}_A-\frac{\lambda}{3}\tilde{\eta}_A^2\right)\epsilon_1.
\end{equation}
Assuming that $\mu$ is almost a constant, we  obtain the power spectrum for the tensor perturbation
\begin{equation}
\label{cosnt0pt}
  \mathcal{P_T}=\frac{2k^3}{2\pi^2}\left|\frac{2u_k}{z_T}\right|^2=\frac{H^2}{\pi}
\frac{1}{(1-\epsilon_1)(1-\delta_1)c_T^3}\left[H_{\mu}^{(1)}\left(\frac{1}{1-\epsilon_1}\frac{c_T k}{aH}\right)\right]^2\left(\frac{c_Tk}{aH}\right)^3.
\end{equation}
On superhorizon scales, $c_Tk\ll aH$, using the asymptotic behavior of the Hankel function, the power spectrum for the tensor perturbation becomes
\begin{equation}\label{cosnt0pt1}
  \mathcal{P_T}=\left.\frac{2^{2\mu}}{(1-\delta_1)c_T^3}\left[\frac{\Gamma(\mu)}{\Gamma(3/2)}\right]^2\left(\frac{H}{2\pi}\right)^2
  \left(1-\epsilon_1\right)^{2\mu-1}\left(\frac{c_Tk}{aH}\right)^{3-2\mu}\right|_{aH=c_T k}.
\end{equation}
The tensor spectral tilt is
\begin{equation}
\label{const0nt}
\begin{split}
n_T&=-2\left(1- \lambda\tilde{\eta}_A-\frac{ \lambda}{3}\tilde{\eta}_A^2\right)\epsilon_1\\
&=-\frac{6\tilde{\eta}_A-6\lambda\tilde{\eta}_A^2-2\lambda\tilde{\eta}_A^3}{3(A+\tilde{\eta}_A)\exp(\tilde{\eta}_A N)-3A}.
\end{split}
\end{equation}
Combining Eqs. \eqref{const0pr1} and \eqref{cosnt0pt1}, we  obtain the tensor-to-scalar ratio
\begin{equation}
\label{const0r}
r=\frac{\mathcal{P_T}}{\mathcal{P_R}}=16(1-\lambda+\lambda\tilde{\eta}_A)\left[ 2^{3-|3+\tilde{\eta}_A|}\times\frac{\Gamma^2(3/2)}{\Gamma^2(|3/2+\tilde{\eta}_A/2|)}\right]\epsilon_1.
\end{equation}
For the constant-roll inflation, with the help of the Gauss-Bonnet coupling, the tensor-to-scalar ratio $r$ is reduced by the factor $(1-\lambda+\lambda\tilde{\eta}_A)$.
However, for the model with large $\tilde{\eta}_A$ like the ultra slow-roll inflation, this reduction does not work.

In the absence of the Gauss-Bonnet coupling, $\lambda=0$, the scalar spectral tilt becomes \cite{Gao:2018cpp}
\begin{equation}
\label{const0ns0}
\begin{split}
n_s-1&=3-|3+\tilde{\eta}_A|-\frac{3(2+A)+(5+2A) \tilde{\eta}_A+\tilde{\eta}_A^2}{|\tilde{\eta}_A+3|}\,\epsilon_1\\
&=3-|3+\tilde{\eta}_A|-\frac{3(2+A)\tilde{\eta}_A+(5+2A) \tilde{\eta}_A^2+\tilde{\eta}_A^3}{|\tilde{\eta}_A+3|[(A+\tilde{\eta}_A)\exp(\tilde{\eta}_A N)-A]}.
\end{split}
\end{equation}
The tensor-to-scalar ratio becomes
\begin{equation}
\label{const0r0}
r=\frac{\mathcal{P_T}}{\mathcal{P_R}}=16\left[ 2^{3-|3+\tilde{\eta}_A|}\times\frac{\Gamma^2(3/2)}{\Gamma^2(|3/2+\tilde{\eta}_A/2|)}\right]\epsilon_1.
\end{equation}
From Eqs. \eqref{const0ns0} and \eqref{const0r0}, for $\tilde{\eta}_A=\alpha$ with $|\alpha|\ll 1$, we get
$n_s-1=-\alpha-(2+A)\epsilon_1$ and $r=16\epsilon_1$. If we choose $\tilde{\eta}_A=-2(3+\alpha)$, then we get
$n_s-1=-2\alpha-(4-3A)\epsilon_1$ and $r=16\epsilon_1$. So the large $\eta$ and small $\eta$ duality \cite{Morse:2018kda} does not hold for $\tilde{\eta}_A$,
but the constant-roll model with large $\tilde{\eta}_A$ still predicts almost scale invariant spectrum.

With the Gauss-Bonnet coupling, the large $\eta$ and small $\eta$ duality \cite{Morse:2018kda} even breaks for the tensor-to-scalar ratio $r$.

\subsection{The model with constant $\epsilon_2$}

For $A=0$, the model with constant $\tilde{\eta}_0$ is the constant-roll inflation with $\epsilon_2$ being a constant.
In this case the scalar spectral tilt is
\begin{equation}
\label{const1ns}
  n_s-1=3-|3+\tilde{\eta}_0|-2\nu_{0} \epsilon_1,
\end{equation}
where
\begin{equation}
\label{const1nu1}
  \nu_{0}=\frac{6(1-\lambda)+(5+4\lambda+3\lambda^2)\tilde{\eta}_0
  +(1+6\lambda+8\lambda^2)\tilde{\eta}_0^2
  +(\lambda+4\lambda^2)\tilde{\eta}_0^3}
  {2|\tilde{\eta}_0+3|(1-\lambda+\lambda\tilde{\eta}_0)},
\end{equation}
and
\begin{equation}
\label{constinu1e}
 \epsilon_1=\exp(-\eta_0 N).
\end{equation}
The tensor-to-scalar ratio is
\begin{equation}
\label{const1r}
 r=16(1-\lambda+\lambda\tilde{\eta}_0)\left[ 2^{3-|3+\tilde{\eta}_0|}\frac{\Gamma^2(3/2)}{\Gamma^2(|3/2+\tilde{\eta}_0/2|)}
 \right]\epsilon_1.
\end{equation}
For the canonical case with $\lambda=0$, we have \cite{Gao:2018cpp}
\begin{gather}
 \label{const1nu10}
n_s-1= 3-|3+\tilde{\eta}_0|-\frac{6+5 \tilde{\eta}_0+\tilde{\eta}_0^2}{|\tilde{\eta}_0+3|}\,\epsilon_1,\\
 \label{const1r10}
  r=16\left[2^{3-|3+\tilde{\eta}_0|}\frac{\Gamma^2(3/2)}{\Gamma^2(|3/2+\tilde{\eta}_0/2|)}\right]\epsilon_1.
\end{gather}
From Eqs. \eqref{const1nu10} and \eqref{const1r10}, for $\tilde{\eta}_0=\alpha$ with $|\alpha|\ll 1$, we get
$n_s-1=-\alpha-2\epsilon_1$ and $r=16\epsilon_1$. If we choose $\tilde{\eta}_0=-2(3+\alpha)$, then we get
$n_s-1=-2\alpha-4\epsilon_1$ and $r=16\epsilon_1$.

\begin{figure}[htbp]
\centering
\includegraphics[width=0.45\textwidth]{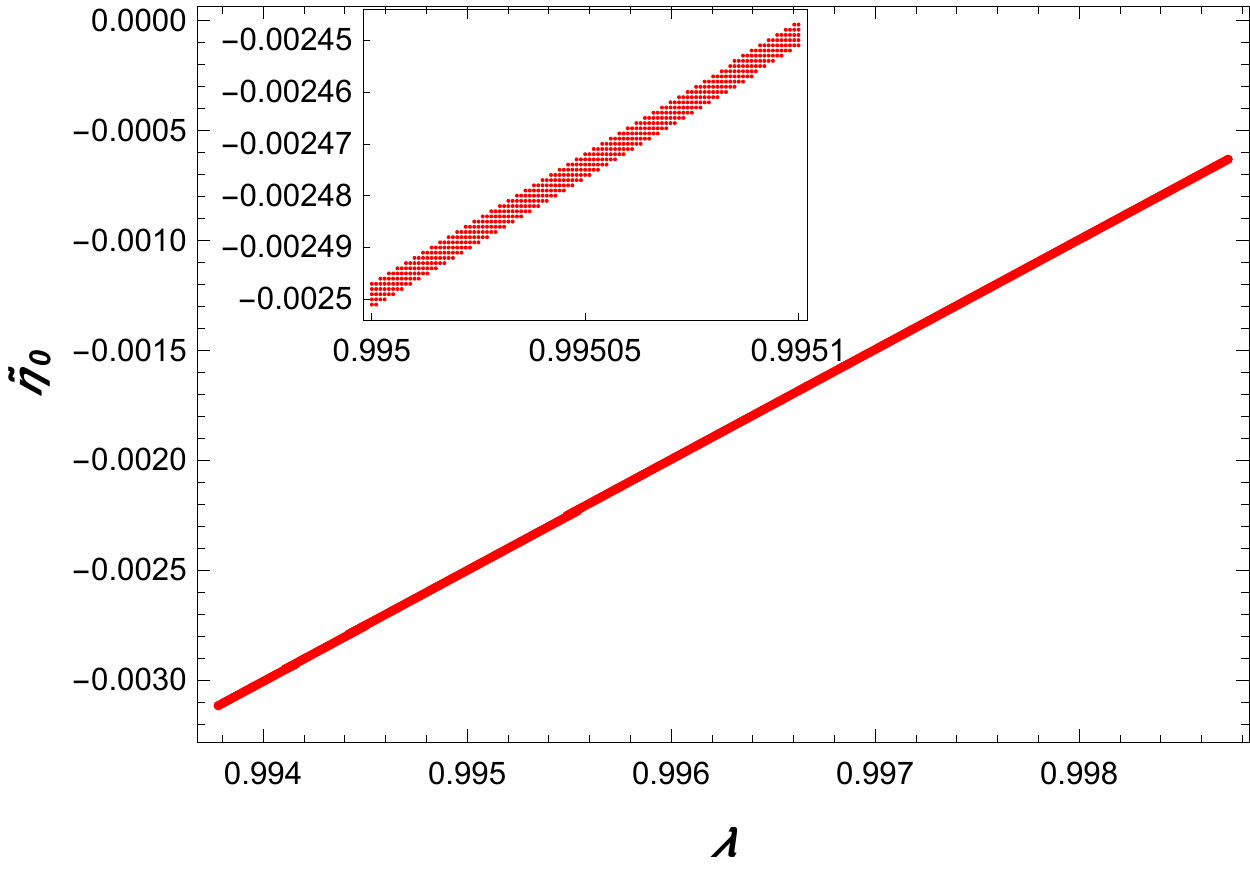}
\caption{The $1\sigma$ constraint on $\lambda$ and $\tilde{\eta}_0$. The inset is a close up of the constraint. The $2\sigma$ and $3\sigma$ constraints are similar and are not shown in the figure.}
\label{p0ab}
\end{figure}

Comparing the predictions from Eqs. \eqref{const1ns} and \eqref{const1r} with the observations \cite{Ade:2015lrj,Array:2015xqh},
we obtain the constraints on the parameters $\tilde{\eta}_0$ and $\lambda$ as shown in Fig. \ref{p0ab}.
Although the observations rule out the canonical model without the Gauss-Bonnet coupling \cite{Gao:2018cpp},
the model with the Gauss-Bonnet coupling is consistent with the observations at the $1\sigma$ C.L. if $\lambda$ is large enough.
From Fig. \ref{p0ab}, we see that the parameters $\tilde{\eta}_0$ and $\lambda$ are highly correlated.

\subsection{The model with constant $\tilde{\eta}_1$}

\begin{figure}[htbp]
\centering
\includegraphics[width=0.45\textwidth]{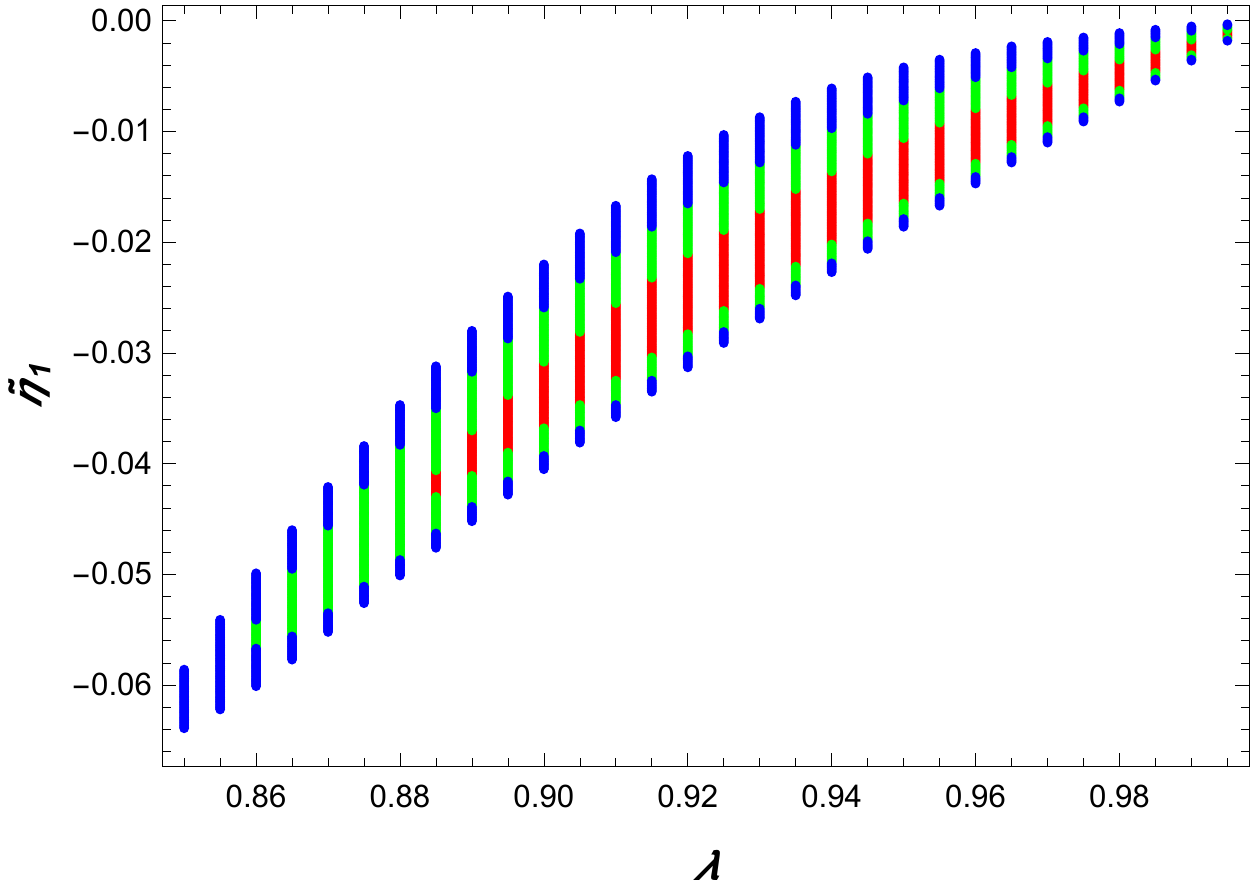}
\caption{The $1\sigma$, $2\sigma$ and $3\sigma$ constraints on $\lambda$ and $\tilde{\eta}_1$.
The red, green and blue regions correspond to the $1\sigma$, $2\sigma$ and $3\sigma$ C.L., respectively.}
\label{p1ab}
\end{figure}

In this case, $A=1$ and we get $\tilde{\eta}_1=\epsilon_2-\epsilon_1$.
For the model with constant $\tilde{\eta}_1$, Eqs. \eqref{eetaA} and \eqref{hj} become
\begin{gather}
\label{eetaA1}
  \tilde{\eta}_1=2\beta_1\left(\frac{H_{,\phi}}{H}\right)^2-\frac{4\beta_1 H_{,\phi\phi}}{H},\\
\label{hj1}
  V(\phi)=3H^2-(2\beta_1^2+12\lambda\beta_1)H_{,\phi}^2,
\end{gather}
where $\beta_{1}=1+\lambda(\tilde{\eta}_1-1)$.
By using Eqs. \eqref{eetaA1} and \eqref{hj1}, we get the potential
\begin{equation}
\label{V1phi1}
V(\phi)=\left\{
\begin{aligned}
 &V_{0}\cosh^4\left[\sqrt{\gamma_1}(\phi-\phi_0)\right]\left(1+
 V_{1}\tanh^2\left[\sqrt{\gamma_1}(\phi-\phi_0)\right]\right),\quad\gamma_1>0,\\
&V_{0}\cos^4\left[\sqrt{-\gamma_1}(\phi-\phi_0)\right]\left(1-
V_{1}\tan^2\left[\sqrt{-\gamma_1}(\phi-\phi_0)\right]\right),\quad \gamma_1<0,
\end{aligned}
\right.
\end{equation}
where
\begin{equation} \gamma_{1}=-\frac{\tilde{\eta}_1}{8\beta_1},\quad V_{1}=\frac{(1+5\lambda+\tilde{\eta}_1\lambda)\tilde{\eta}_1}{3}.
\end{equation}
The scalar spectral tilt  is
\begin{equation}\label{const2ns}
n_s-1=3-|3+\tilde{\eta}_1|-2\nu_1\epsilon_1,
\end{equation}
where
\begin{equation}\label{const2nu1}
    \nu_{1}=\frac{9(1-\lambda)+(7+8\lambda+3\lambda^2)\tilde{\eta}_1
    +(1+10\lambda+8\lambda^2)\tilde{\eta}_1^2
  +(\lambda+4\lambda^2)\tilde{\eta}_1^3}
  {2|3+\tilde{\eta}_1|(1-\lambda+\lambda\tilde{\eta}_1)},
\end{equation}
and
\begin{equation}
\label{const2nu1e}
 \epsilon_1=\frac{\tilde{\eta}_1}{(1+\tilde{\eta}_1)\exp(\tilde{\eta}_1 N)-1}.
\end{equation}
The tensor-to-scalar ratio is
\begin{equation}\label{const2r}
r=16(1-\lambda+\lambda\tilde{\eta}_1)\left[ 2^{3-|3+\tilde{\eta}_1|}\frac{\Gamma^2(3/2)}{\Gamma^2(|3/2+\tilde{\eta}_1/2|)}
 \right]\epsilon_1.
\end{equation}
Comparing the predictions from Eqs. \eqref{const2ns} and \eqref{const2r} with the observations \cite{Ade:2015lrj,Array:2015xqh},
we obtain the constraints on the parameters $\tilde{\eta}_1$ and $\lambda$ as shown in Fig. \ref{p1ab}.
Since $\lambda=0$ is ruled out, so
the observations rule out the model without the Gauss-Bonnet coupling.
With the Gauss-Bonnet coupling, the model is consistent with the observations at the $1\sigma$ C.L.

\subsection{The model with constant $\eta_H$}

In this case, $A=2$ and $\tilde{\eta}_2=-2\eta_H$,
so constant $\tilde{\eta}_2$ is the constant-roll inflation with $\eta_H$ being a constant.
The scalar spectral tilt is
\begin{equation}
\label{const3ns}
n_s-1=3-|3+\tilde{\eta}_2|-2\nu_2\epsilon_1,
\end{equation}
where
\begin{equation}
\label{const3nu1}
  \nu_{2}=\frac{12(1-\lambda)+(9+12\lambda+3\lambda^2)\tilde{\eta}_2
  +(1+14\lambda+8\lambda^2)\tilde{\eta}_2^2
  +(\lambda+4\lambda^2)\tilde{\eta}_2^3}
  {2|\tilde{\eta}_2+3|(1-\lambda+\lambda\tilde{\eta}_2)},
\end{equation}
and
\begin{equation}
\label{constinu3e}
 \epsilon_1=\frac{\tilde{\eta}_2}{(2+\tilde{\eta}_2)
  \exp(\tilde{\eta}_2 N)-2}.
\end{equation}
The tensor-to-scalar ratio is
\begin{equation}\label{const3r}
r=16(1-\lambda+\lambda\tilde{\eta}_2)\left[ 2^{3-|3+\tilde{\eta}_2|}\frac{\Gamma^2(3/2)}{\Gamma^2(|3/2+\tilde{\eta}_2/2|)}
 \right]\epsilon_1.
\end{equation}
For the canonical case with $\lambda=0$, we have
\begin{gather}
 \label{const3nu10}
 n_s-1= 3-|3+\tilde{\eta}_2|-\frac{12+9 \tilde{\eta}_2+\tilde{\eta}_2^2}{|\tilde{\eta}_2+3|}\times\epsilon_1,\\
 \label{const3r10}
  r=16\left[2^{3-|3+\tilde{\eta}_2|}\frac{\Gamma^2(3/2)}{\Gamma^2(|3/2+\tilde{\eta}_2/2|)}\right]\times\epsilon_1.
\end{gather}
For $\tilde{\eta}_2=\alpha$ with $|\alpha|\ll 1$, we get
$n_s-1=-\alpha-4\epsilon_1$ and $r=16\epsilon_1$. If we choose $\tilde{\eta}_2=-2(3+\alpha)$, then we get
$n_s-1=-2\alpha+2\epsilon_1$ and $r=16\epsilon_1$.

\begin{figure}[htbp]
\centering
\includegraphics[width=0.45\textwidth]{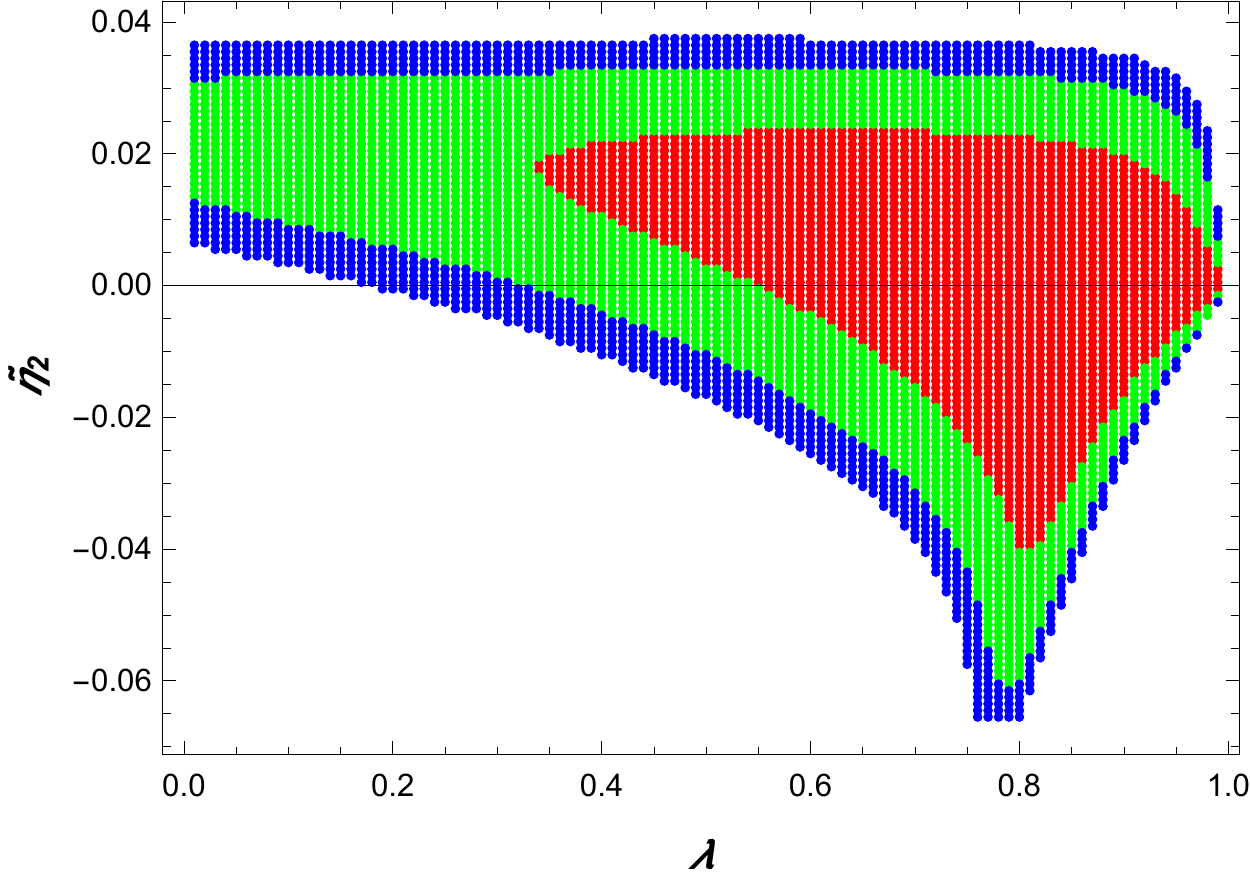}
\caption{The $1\sigma$, $2\sigma$ and $3\sigma$ constraints on $\lambda$ and $\tilde{\eta}_2$.
The red, green and blue regions correspond to the $1\sigma$, $2\sigma$ and $3\sigma$ C.L., respectively.}
\label{p4ab}
\end{figure}

Comparing the predictions from Eqs. \eqref{const3ns} and \eqref{const3r} with the observations \cite{Ade:2015lrj,Array:2015xqh},
we obtain the constraints on the parameters $\tilde{\eta}_2=-2\eta_H$ and $\lambda$ as shown in Fig. \ref{p4ab}.
For the constant $\eta_H$ inflation without the Gauss-Bonnet coupling,
the predictions are consistent with the observations at the $2\sigma$ C.L. \cite{Yi:2017mxs}.
With the Gauss-Bonnet coupling, the predictions are consistent with the observations at the $1\sigma$ C.L.
If we take $\lambda=0.81$, $\tilde{\eta}_2=-0.01$ ($\eta_H=0.005$), we get $n_s=0.968$ and $r=0.03$.
Since the observations require that $\epsilon_1$ and $\tilde{\eta}_2$  are both small,
so the slow-roll conditions are satisfied and the constant-roll inflation with constant $\eta_H$  is also a slow-roll inflation.

\subsection{The model with constant $\eta_V$}

Finally, we consider the case $A=4$, the model with constant $\tilde{\eta}_4$ which includes the slow-roll inflation with $\eta_V$ being a constant\cite{Gao:2017owg}.
The scalar spectral tilt $n_s$ and the tensor-to-scalar ratio $r$ are
\begin{gather}
\label{const4ns}
n_s-1=3-|3+\tilde{\eta}_4|-2\nu_4\epsilon_1,\\
\label{const4r}
 r=16(1-\lambda+\lambda\tilde{\eta}_4)\left[ 2^{3-|3+\tilde{\eta}_4|}\frac{\Gamma^2(3/2)}{\Gamma^2(|3/2+\tilde{\eta}_4/2|)}
 \right]\epsilon_1.
\end{gather}
where
\begin{equation}\label{const4nu1}
  \nu_{4}=\frac{18(1-\lambda)+(13+20\lambda+3\lambda^2)\tilde{\eta}_4
  +(1+22\lambda+8\lambda^2)\tilde{\eta}_4^2
  +(\lambda+4\lambda^2)\tilde{\eta}_4^3}
  {2|\tilde{\eta}_4+3|(1-\lambda+\lambda\tilde{\eta}_4)}.
\end{equation}
For the canonical case with $\lambda=0$, we have \cite{Gao:2018cpp}
\begin{gather}
 \label{const4nu10}
n_s-1= 3-|3+\tilde{\eta}_4|-\frac{18+13 \tilde{\eta}_4+\tilde{\eta}_4^2}{|\tilde{\eta}_4+3|}\,\epsilon_1,\\
 \label{const4r10}
 r=16\left[2^{3-|3+\tilde{\eta}_4|}\frac{\Gamma^2(3/2)}{\Gamma^2(|3/2+\tilde{\eta}_4/2|)}
 \right]\epsilon_1.
\end{gather}
For $\tilde{\eta}_4=\alpha$ with $|\alpha|\ll 1$, we get
$n_s-1=-\alpha-6\epsilon_1$ and $r=16\epsilon_1$. If we choose $\tilde{\eta}_4=-2(3+\alpha)$, then we get
$n_s-1=-2\alpha+8\epsilon_1$ and $r=16\epsilon_1$.

\begin{figure}[htbp]
\centering
\includegraphics[width=0.45\textwidth]{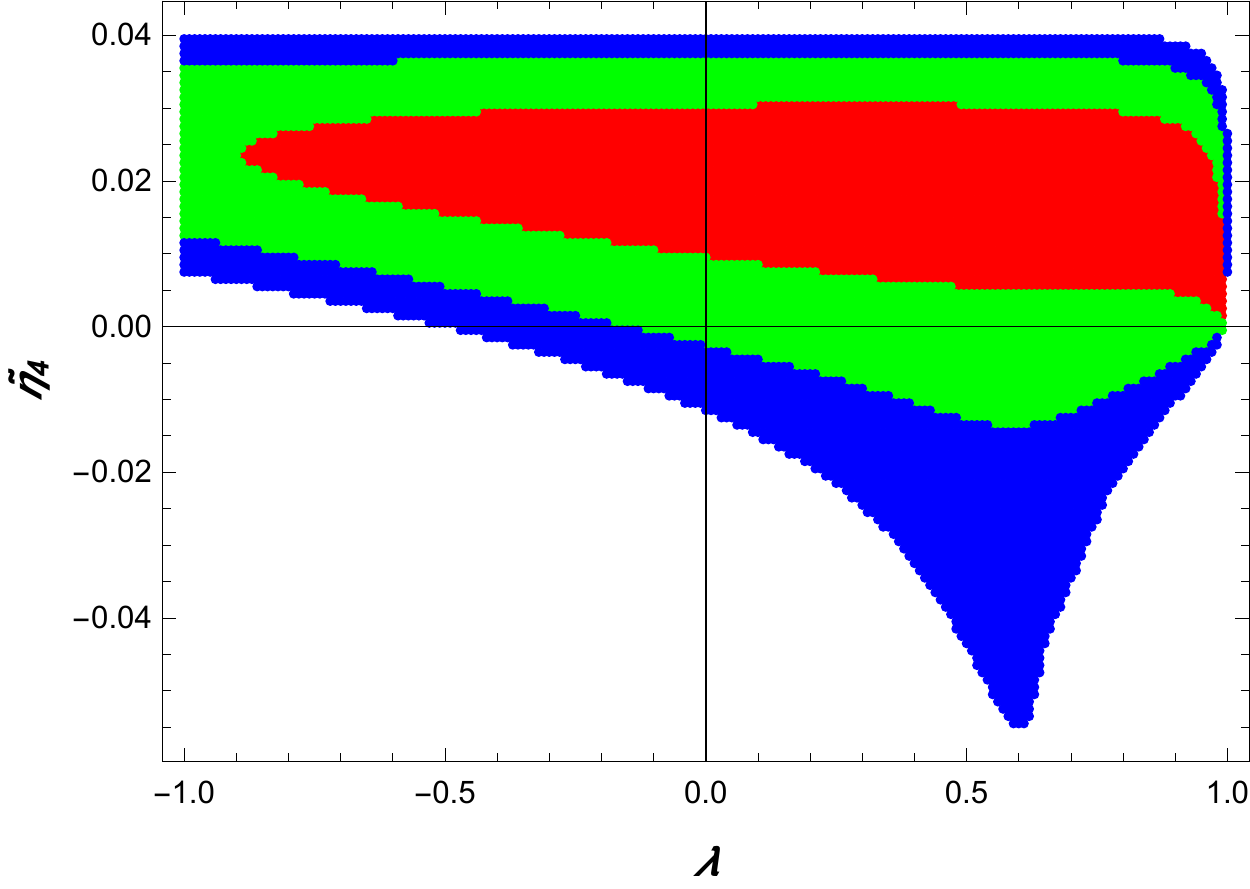}
\caption{The $1\sigma$, $2\sigma$ and $3\sigma$ constraints on $\lambda$ and $\tilde{\eta}_4$.
The red, green and blue regions correspond to the $1\sigma$, $2\sigma$ and $3\sigma$ C.L., respectively.}
\label{p5ab}
\end{figure}

Comparing the predictions from Eqs. \eqref{const4ns} and \eqref{const4r} with the observations \cite{Ade:2015lrj,Array:2015xqh},
we obtain the constraints on the parameters $\tilde{\eta}_4$ and $\lambda$ as shown in Fig. \ref{p5ab}.
This model is consistent with the observations at the $1\sigma$ C.L.

\subsection{The potentials for small $\tilde{\eta}_A$}

From the above discussions, we see that the observations require that $\tilde{\eta}_A$ is small. In this subsection,
we consider the potentials for the constant-roll inflation with small $\tilde{\eta}_A$.
To the first order of approximation, Eqs. \eqref{eetaA} and \eqref{hj} become
\begin{gather}
\label{eetaAsl}
  \tilde{\eta}_A \approx 2(2-A)\bar\lambda\left(\frac{H_{,\phi}}{H}\right)^2-\frac{4\bar\lambda H_{,\phi\phi}}{H},\\
 \label{hjsl}
 V(\phi)\approx 3H^2-(12\lambda\bar\lambda+2\bar\lambda^2)H_{,\phi}^2,
\end{gather}
where $\bar\lambda=1-\lambda$.
For the slow-roll case, we can derive the potential for the model with
constant $\tilde{\eta}_A$ by using Eqs. \eqref{eetaAsl} and \eqref{hjsl}.
For convenience, we introduce the function $X(\phi)$,
\begin{equation}\label{xfun}
  X(\phi)=H(\phi)^{\frac A 2}.
\end{equation}
Substituting the function $X(\phi)$ into Eq. \eqref{eetaAsl}, we get
\begin{equation}\label{xequ}
  \frac{X_{,\phi\phi}}{X}=\gamma_A,
\end{equation}
where $\gamma_A=-A\tilde{\eta}_A/(8\bar\lambda)$. The solution to Eq. \eqref{xequ} is
\begin{equation}
\label{xs1}
  X(\phi)=c_1 \exp[\sqrt{\gamma_A}(\phi-\phi_0]+c_2 \exp[-\sqrt{\gamma_A}(\phi-\phi_0],
\end{equation}
if $\gamma_A>0$, where $c_1$ and $c_2$ are  integration constants. For any values of $c_1$ and $c_2$,
we can choose the value of $\phi_0$ so that the solution falls into one of the following
three classes
\begin{gather}
\label{exp1}
\text{(1)} \quad X(\phi)=M\exp(\pm\sqrt{\gamma_A}\phi),\quad c_1c_2=0,\\
\label{sinh1}
\text{(2)} \quad X(\phi)=M\sinh\left(\sqrt{\gamma_A}\phi\right),\quad c_1c_2<0,\\
\label{cosh1}
\text{(3)} \quad X(\phi)=M \cosh\left(\sqrt{\gamma_A} \phi \right),\quad c_1c_2>0,
\end{gather}
where $M>0$.
The potential for the  case (1) is
\begin{equation}
 \label{expA}
V(\phi)=V_0 \exp\left(\pm\frac{4 \sqrt{\gamma_A}}{A}\phi\right),
\end{equation}
the  potential for the case (2) is
\begin{equation}
 \label{sinhA}
V(\phi)=V_0 \sinh^{4/A} ( \sqrt{\gamma_A} \phi)
\left[1+V_{A}\coth^2( \sqrt{\gamma_A} \phi)\right],
\end{equation}
and the  potential for the case  (3)  is
\begin{equation}
 \label{coshA}
 V(\phi)=V_0 \cosh^{4/A} ( \sqrt{\gamma_A} \phi )
\left[ 1+V_{A} \tanh^2( \sqrt{\gamma_A} \phi )\right],
\end{equation}
where $V_{A}=(5\lambda+1)\tilde{\eta}_A/(3A)$.

\begin{figure}[htbp]
\centering
\includegraphics[width=0.55\textwidth]{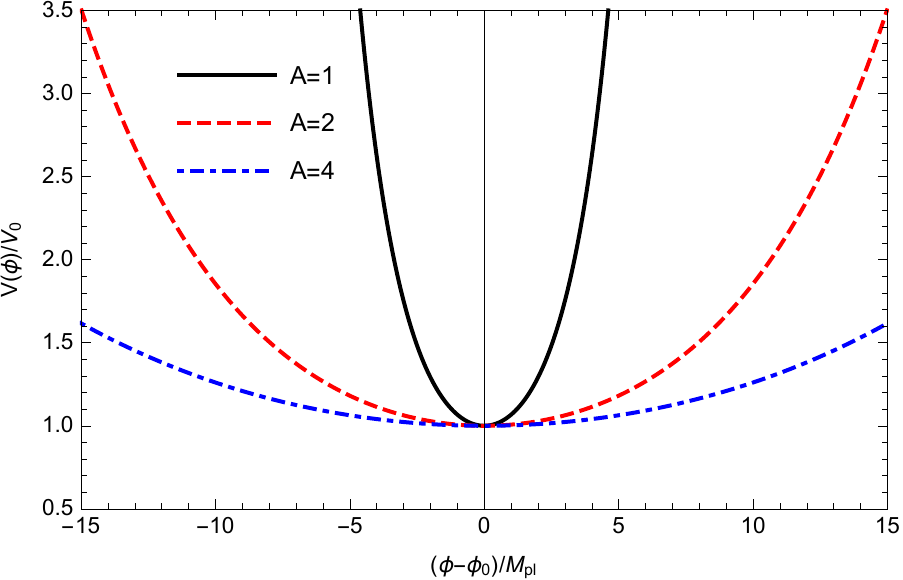}
\caption{The potentials \eqref{coshA} normalized by $V_0$ for $A=1$, $A=2$ and $A=4$.}
\label{p7}
\end{figure}

For $\gamma_A<0$, the solution to Eq. \eqref{xequ} is
\begin{equation}
 \label{xs2}
 X(\phi)=M \cos[\sqrt{-\gamma_A}(\phi-\phi_0],
\end{equation}
and the potential is
\begin{equation}
\label{cosA}
V(\phi)=V_{0}\cos^{4/A}\left[ \sqrt{-\gamma_A}(\phi-\phi_0)\right]\left(1-
  V_{A}\tan^2\left[\sqrt{-\gamma_A}(\phi-\phi_0)\right]\right).
\end{equation}
Note that if $A=1$, Eqs. \eqref{coshA} and \eqref{cosA} give the potential \eqref{V1phi1} for small $\tilde{\eta}_1$.
The potentials \eqref{coshA} for $A=1$, $A=2$ and $A=4$ are shown in Fig. \ref{p7}.

\section{Conclusion}
\label{sec-5}

For the slow-roll inflation, the reciprocal relation $\xi=3\lambda/(4V)$ can be derived from the condition $\delta_1=2\lambda\epsilon_1$.
To overcome the reheating problem due to the divergence of the coupling, we take the coupling $\xi=3\lambda/(4V+\Lambda_0)$ instead and use
the reciprocal relation $\xi=3\lambda/(4V)$ as the approximation during the slow-roll period.
With the help of the Gauss-Bonnet coupling and the condition $\delta_1=2\lambda\epsilon_1$,
the tensor-to-scalar ratio $r$ is reduced by a factor of $1-\lambda$ so that the results become more favorable by the observations.
Therefore, inflation models with large $r$ can be saved by the Gauss-Bonnet coupling.
For the model with large $r$, such as the natural inflation ruled out by the observations at the $1\sigma$ confidence level,
we find that if $\lambda>0.55$, it will be consistent with the observations at the $1\sigma$ confidence level.

We use a general parametrization $\tilde{\eta}_A=\epsilon_2-A\epsilon_1$ to discuss
different constant-roll inflations with the condition \eqref{condition1}.
For the constant-roll inflation, the tensor-to-scalar ratio $r$ is reduced by a factor of $1-\lambda+\lambda\tilde\eta_A$,
so the reduction does not work for the models with large $\tilde\eta_A$ like the ultra slow-roll inflation.
For the model with constant $\tilde{\eta}_A$, we derive the formulae for the power spectra of both the scalar and tensor perturbations.
The formulae are applied to four specific models and the observational data are used to constrain the model parameters.
For the case $A=0$, we have $\tilde{\eta}_0=\epsilon_2$ and this corresponds to
the constant-roll inflation with constant $\epsilon_2$. This model is consistent with the observations if $\lambda>0.99$.
If $A=1$, we have $\tilde{\eta}_1=\epsilon_2-\epsilon_1$, and
the model with constant $\tilde{\eta}_1$ is consistent with the observations if $\lambda>0.84$.
Without taking the slow-roll approximation, the potential for the model is derived.
For the case $A=2$, we have $\tilde{\eta}_2=-2\eta_H$ and this corresponds to
the constant-roll inflation with constant $\eta_H$.
The constraints on the model parameters $\tilde{\eta}_2$ and $\lambda$ are obtained.
For the case $A=4$, in the slow-roll approximation, constant $\tilde{\eta}_4$ corresponds to the constant-roll
inflation with constant $\eta_V$. The model is consistent with the observations even when the Gauss-Bonnet coupling is absent.
For the models with constant $\tilde{\eta}_0$, $\tilde{\eta}_1$, $\tilde{\eta}_2$ and $\tilde{\eta}_4$,
the observations constrain the model parameter $\tilde{\eta}_A$ to be small, so these constant-roll inflations are
also slow-roll inflations. Using the slow-roll approximation, the potentials for these models are obtained.
In conclusion, the Gauss-Bonnet coupling and the condition $\delta_1=2\lambda\epsilon_1$ help
inflation models to be consistent with the observations.

\begin{acknowledgments}
This work was supported in part by the National Natural Science
Foundation of China under Grant Nos. 11875136 and 11475065 and the Major Program of the National Natural Science Foundation of China under Grant No. 11690021. M. S. would like to thank the Higher Education Commission of Pakistan for a Ph.D. scholarship.
\end{acknowledgments}

%

\end{document}